\documentclass{article}
\usepackage{amssymb}
\newcommand\E{\rm e}
\newcommand\D{\rm d }
\begin{document}
\title{LOCAL FIELD THEORY ON $\kappa$-MINKOWSKI SPACE, 
\\ 
STAR PRODUCTS AND NONCOMMUTATIVE 
TRANSLATIONS\footnote{To be published in Proceedings of 
Colloqium on
Quantum Groups and Integraable   Systems, Prague, June 2000 
(Czech. J. Phys. {\bf 50}, 2000, in press}}
\author{P. Kosi\'{n}ski
 \\ Institute of Physics, 
University of L\'{o}d\'{z}, \\
ul. Pomorska 149/53 90--236 L\'{o}d\'{z}, Poland \\ \\
J. Lukierski \\
Institute of Theoretical Physics, 
 University of Wroclaw \\ pl. M. Borna 9, 50-205 Wroc\l aw, Poland 
\\ \\
P. Ma\'{s}lanka \\
Institute of Physics, 
University of L\'{o}d\'{z}, \\
ul. Pomorska 149/53 90--236 L\'{o}d\'{z}, Poland }
\date{}
%%%%%%%%%%%%%%%%%%%%%%%%%%%%%%%%%%%
%%%%%%%%%%%%%%%%%%%%%
\maketitle

\begin{abstract}
We consider local field theory on $\kappa$-deformed Minkowski
space which is an example of solvable Lie-algebraic
noncommutative structure. Using integration formula over
$\kappa$-Minkowski space and $\kappa$-deformed Fourier 
transform
we consider for deformed local fields the reality conditions as
well as deformation of action functionals  in standard Minkowski
space. We present explicite formulas for two equivalent star
products describing CBH quantization of field theory on
$\kappa$-Minkowski space. We express also via star product
technique the noncommutative translations in $\kappa$-Minkowski
space by commutative translations in standard Minkowski space.
\end{abstract}

\section{Introduction}    

Recently the idea that the commuting space-time coordinates
${x}_{\mu}$ are replaced by algebra of noncommuting
generators $\hat{x}_{\mu}$ became quite popular (see e.g.
[1-5]). The noncommutative Minkowski space on one side 
represents 
algebraically the quantum gravity corrections [1,2], on other hand
describes the end points of strings coupled to antisymmetric
two-index gauge field $B_{\mu\nu}$ (in $D=10$) [3,5] and boundary
strings for membranes coupled to the antisymmetric three-index
gauge field $C_{\mu\nu\rho}$ (in $D=11$). In general the
noncommutativity in ``new" string theory occurs for the
$(p+1)$ world-volume 
 coordinates of Dirichlet $p$-branes and follows from the
modification of standard quantization rules of the classical
($p+1$)-dimensional $p$-brane fields in the presence of external
generalized gauge potentials.

In most of the papers dealing with noncommutative field theory
one considers the Dopplicher--Fredenhagen--Roberts (DFR)
deformation [1] of space--time coordinates
\begin{equation}
\left[ \hat{x}_{\mu}, \hat{x}_{\nu}\right] = i\theta_{\mu\nu} 
\qquad \theta_{\mu\nu} =- \theta_{\nu\mu} \ \mbox{constant}\, .
 \end{equation}
  In the presence of DFR noncommutativity the relativistic
invariance remains classical and the translations $x_{\mu} \to  
 \hat{x}_{\mu}^{\prime} = \hat{x}_{\mu} + a_{\mu}$ preserving
(1) are commutative, i.e. $[a_{\mu}, a_{\nu}]=0$. This feature is
not valid if  $\theta_{\mu\nu}$ depends on $
\hat{x}_{\mu}$, in particular if we assume  the linear 
$ \hat{x}_{\mu}$--dependence, i.e.
\begin{equation}
\left[ \hat{x}_{\mu}, \hat{x}_{\nu}\right] = i c_{\mu\nu} ^{\lambda}
\hat{x}_{\lambda}\, .
\end{equation}
In such a case the translations 
\begin{equation}
 \hat{x}_{\mu}^{\prime} 
 = \hat{x}_{\mu} + 
\hat{v}_{\mu}\, ,
\end{equation}
are not commutative, 
commute with $\kappa$-Minkowski coordinates ($[\hat{x}_{\mu},
\hat{v}_{\mu}]=0$) 
 and provide second copy of the algebra (2)

\begin{equation}
\left[ \hat{v}_{\mu}, \hat{v}_{\nu}\right] 
= i c_{\mu\nu} ^{\lambda}
\hat{v}_{\lambda}\, .
\end{equation}
One can describe (3) as a coproduct 
$\Delta(\hat{x}_{\mu}) = x_{\mu}\otimes 1  + 1 \otimes x_{\mu}$, 
where
$\hat{x}_{\mu} \otimes 1 \equiv \hat{x}_{\mu}$ and
$\hat{v}_{\mu} = 1\otimes \hat{x}_{\mu}$. It appears that for
particular choices of the structure constants
$c_{\mu\nu}^{\lambda}$ the noncommutative translations (4) can
be incorporated as translation sector of the quantum--deformed
Poincar\'{e} group, with noncommutative group
parameters\footnote{For classification of quantum Poincar\'{e}
groups see [10].}

It should be stressed that a natural mathematical framework
providing noncommutative space-time is given by quantum groups
describing the deformations of space-time symmetries. The
formalism of quantum groups (see e.g. [9]) provides the modified 
 symmetries in
the form of a dual pair of Hopf algebras $A\otimes A^{\star}$
(in the applications discussed here $A$ describes the deformation
 of Poincar\'{e} group $A$ and
$A^{\star}$ - the deformation of Poincar\'{e} algebra) where the
 noncommutative coordinates ${\hat{x}}_{\mu}\in A$ and the 
fourmomenta
${\hat{p}}_{\mu}\in A^{\star}$. In particular
 coproducts of ${\hat{x}}_{\mu}$ and ${\hat{p}}_{\mu}$ describe
 the addition laws of noncommutative coordinates and
noncommutative momenta. 

In this talk
 we
 will consider a particular example of
so-called $\kappa$-deformation of space-time symmetries [11--15]
with noncommutativity of space-time coordinates described by
the  following solvable Lie-algebraic  relations:
\begin{equation}
\left[ {\hat{x}}_{0}, {\hat{x}}_{i} \right] = 
\frac{i}{ \kappa} {\hat{x}}_{i}\, ,
\qquad
\left[ {\hat{x}}_{l}, {\hat{x}}_{j} \right] = 0\, .
\end{equation}
The primitive coproducts lead to the
   addition law (3) for the
noncommutative coordinates 
and the dual fourmomentum generators are described by 
\underline{commuting} fourmomenta $p_{\mu}$, with 
  coproducts generating the following nonabelian 
 addition law 
\begin{equation}
p_{0}^{(1+2)} = p_{0}^{(1)} + p_{0}^{(2)}\, ,
\qquad
p_{i}^{(1+2)}= p_{i}^{(1)}\E^{-{p_{0}^{(2)}\over \kappa}} +
p_{i}^{(2)}\, .
\end{equation}

In this lecture  we shall describe some new aspects of 
 the deformation of local
relativistic $D=4$ field theory on noncommutative space-time
(5), with taking into consideration the relations (6). 
In particular we shall consider new $\kappa$--deformed Fourier
transform of local fields on $D=4$ $\kappa$--Minkowski space,
the reality conditions and action integrals for
complex $\kappa$--deformed local fields. Our next result here is
the explicite CBH formula for star-product quantization of
$\kappa$--deformed field theory\footnote{For general
Campbell--Baker-Hausdorff quantization formula describing
quantization of linear Poisson brackets see e.g. [16,17].} 
Also we shall show that the noncommutative translations (3) of
field arguments can be expressed via 
 star product technique in terms of classical, commutating
translation parameters.

It should be recalled that the $\kappa$--deformed local
interaction vertices lead to modified Feynman diagram rules with
deformed fourmomentum conservation laws [18,19]. It is
interesting to mention that the effects of nonconservation of
momenta has been observed recently also in the noncommutative
field theory constructed
  without any reference to Hopf--algebraic structures
(see e.g. [20]).

\section{$\kappa$-Deformed Fourier Transform and Reality 
Conditions}
The duality structure provides natural definition of Fourier
transform mapping the functions (fields) on noncommutative space-
time
$\hat{x}_{\mu}$ into the functions of dual fourmomentum space [9].

In $\kappa$-deformed Minkowski space one can write the following 
formulae\footnote{For the formulae in $1+1$ Minkowski space see 
[21]. It
  should be mentioned that we introduce here modified
formula (7) in comparison with the one presented in [18]. In
the formulae (7--8) the Fourier transform (8) is inverse to
the  one given by (7).}

\begin{equation}
\Phi(\hat{x}) =
{1\over (2\pi)^{2}} \int \D^4 p\  \widetilde{\Phi}_{\kappa}
(p)\, : \E^{ip\hat{x}}:\, = \widehat{F}_{\kappa}
\widetilde{\Phi}(p)\, ,
\end{equation}

\begin{equation}
{\widetilde{\Phi}} (p) =
{1\over (2\pi)^{2}} 
 \int\!\!\!\int \D^4\  {\hat{x}}
 {\Phi} ({\hat{x}} ) 
\, : \E^{-ip\hat{x}}:\, = \widehat{F}_{\kappa}^{-1}
{\Phi}({\hat{x}})\, ,
\end{equation}
where $p\hat{x}=\vec{p}\vec{x} - p_{0}x_{0}$ and
\begin{equation}
 : \E^{ip\hat{x}}: =   \E^{- ip_{0}\hat{x}_{0}}
   \E^{i\vec{p}\vec{x}}\, ,
\end{equation}

\begin{equation}
\widetilde{\Phi}_{\kappa}(p)  =   \E^{ {3p_{0}\over \kappa}}
{\widetilde{\Phi}}\, \left( 
  \E^{{p_{0}\over \kappa}}\vec{p}, p_{0}\right)\, .
\end{equation}
It should be pointed out that the $\kappa$-deformed exponential 
has
been chosen in the way providing the $\kappa$-deformed addition
rule of fourmomenta  (see also (6))
\begin{equation}
 : \E^{ip^{(1)}\hat{x}}: \, :  
  \E^{ip^{(2)}\hat{x}}: \,  =
 :  \E^{ip^{(1+2)}\hat{x}}: 
 \end{equation}
and the $\kappa$-deformed integration over
$\kappa$-Minkowski space is generated by the formula

\begin{equation}
{1\over (2\pi)^{4}}
\int\!\!\!\int \D^{4}\ \hat{x}\, : \E^{ip\hat{x}}: \, =
\delta^{4} (p)\, .
\end{equation}

The formulae (11-12) imply that

\begin{equation}
{1\over (2\pi)^{2}}
\int\!\!\!\int \D^{4}\  {\hat{x}} \, \Phi({\hat{x}}) =
{\widetilde{\Phi}}_{\kappa} (0) = {\widetilde{\Phi}} (0)\, ,
\end{equation}

\begin{eqnarray}
\int\!\!\!\int \D^{4}\  \hat{x} \, \Phi^{2}(\hat{x}) &= &
\int \D^{4} p \ \D^{4} q \
\widetilde{\Phi} \left( \E^{{p_{0}\over \kappa}} \vec{p}, p_{0}\right)
\widetilde{\Phi} \left( \E^{{q_{0}\over \kappa}} \vec{q}, q_{0}\right)
\cr\cr
& &\,\cdot \, \delta\left(p_{0}+q_{0}\right)  \delta^{(3)}
\left(\vec{p} \E^{-{q_{0}\over \kappa}} + \vec{q} \right)
\cr\cr
& = &
\int \D^{4} p\
\widetilde{\Phi} \left( 
-  \vec{p}, - p_{0}\right)
\widetilde{\Phi}\left( \E^{{p_{0}\over \kappa}}  \vec{p}
,p_{0}\right)\, .
\end{eqnarray}

Let us introduce now the Hermitean conjugation of the field (7). 
Defining
\begin{eqnarray}
 \Phi^{+}(\hat{x}) &=
&
 {1\over (2\pi)^{2}} 
\int \D^{4} p\  \widetilde{\Phi}\ ^{*}_{\kappa}(p)
 \E^{- i{\vec{p}}{\vec{x}} }
 \E^{i p_{0} \hat{x}_{0}} 
 \cr
 &=&
   {1\over (2\pi)^{2}} 
 \int \D^{4} p \ \widetilde{\Phi}\ ^{+}_{\kappa}(p):
 \E^{i{{p}}{\hat{x}} }:
\end{eqnarray}
and using the formula
\begin{equation}
 \E^{- i{\vec{p}}{\vec{x}} }
 \E^{i p_{0} {x}_{0} } =
\E^{ip_{0} \hat{x}_{0}} 
 \E^{-i e^{p_{0}\over \kappa} \vec{p} \vec{x}}\, ,
 \end{equation}
 one gets
 \begin{equation}
  \Phi^{+}_{\kappa} (p) 
  =
    \widetilde{\Phi}\ ^{\star}(-p) \, .
\end{equation}
Let us observe that the reality condition 
$  \Phi^{+}(\hat{x}) = 
  \Phi(\hat{x})$ gives
  \begin{equation}
    \Phi_{\kappa}(p) =
      \Phi^{+}_{\kappa}(p) \leftrightarrow
    \Phi_{\kappa}(p) =
     \widetilde{ \Phi}\ ^{\star}(-p) \, .
     \end{equation}
     For real noncommutative fields one gets, using (18)

\begin{eqnarray}
\int\!\!\!\int \D^{4}\  \hat{x} \, \Phi^{2}(\hat{x}) &= &
\int \D^{4} p\  
 \E^{{3 p_{0}\over \kappa}} 
\, \widetilde{\Phi} (p)
\, \widetilde{\Phi}\ ^{\star}(p)
\cr\cr
&=&
\int \D^{4} p  \
 \E^{-{ 3 p_{0}\over \kappa}} 
\widetilde{\Phi}^{\star}_{\kappa} (p)
\widetilde{\Phi}\ _{\kappa}(p)\, .
     \end{eqnarray}
     
For complex noncommutative fields one can write
\renewcommand{\theequation}{20\alph{equation}}
\setcounter{equation}{0}
\begin{equation}
\int\!\!\!\int \D^{4}\  \hat{x} \, \Phi(\hat{x})  \Phi^{+}(\hat{x}) 
=
\int\D^{4} p\
 \widetilde{ \Phi}(p) 
    \widetilde{\Phi}\ ^{\star}(p)\, ,
\end{equation}

\begin{equation}
\int\!\!\!\int \D^{4}\  \hat{x} \, \Phi^{+}(\hat{x})  \Phi(\hat{x}) 
     =
     \int\D^{4} p\
e^{3p_{0}\over \kappa} 
    \widetilde{\Phi}\ ^{\star}(p)
 \widetilde{ \Phi}(p) \, .
 \end{equation}
 \renewcommand{\theequation}{\arabic{equation}}
\setcounter{equation}{20}
     It should be added that we obtain alternative definition of
Fourier transform if in the formula (8) the $\kappa$--deformed
exponential is located on left side of the field $ \Phi(\hat{x})$.

\section{Star Product for $\kappa$--Deformed Field Theory}

It is known [16,17,4,22] that the multiplication of two group     
 elements with Lie algebra generators $\hat{x}_{\mu}$ described
by relation (2) is given by CBH formula
\begin{equation}
e^{i \alpha^{\mu}\hat{x}_{\mu}} \, 
e^{i \beta^{\mu}\hat{x}_{\mu} } =
e^{i \gamma ^{\mu} (\alpha,\beta) \hat{x}_{\mu}}\, ,
     \end{equation}
     where
\begin{eqnarray}
 \gamma^{\mu} (\alpha,\beta) &=&
 \alpha^{\mu} + \beta^{\mu} + c^{\mu}_{\rho\tau}
 \alpha^{\rho}\beta^{\tau} 
 \cr\cr
&&  + \ {1\over 12} \, c^{\mu}_{\rho\tau}      c^{\rho}_{\lambda\nu}
     \left(
     \alpha^{\tau}\alpha^{\lambda} \beta^{\nu} +
     \beta^{\tau} \beta^{\lambda} \alpha^{\nu} \right) +
\ldots\, .
     \end{eqnarray}
     The noncommutative group multiplication (21) is translated
into the framework of star products by substitution 
$\hat{x}_{\mu} \to x_{\mu}$, where $x_{\mu}$ are classical
Minkowski coordinates and copying the relation (21)

\begin{equation}
e^{i \alpha^{\mu} {x}_{\mu} } \, \star 
e^{i \beta ^{\mu}{x}_{\mu} } =
e^{i \gamma ^{\mu} (\alpha,\beta) {x}_{\mu} }\, .
     \end{equation}     
The formula ( 23) describes the CBH star product and the
associativity of group multiplication leads to the associativity
of CBH star--multiplication. From relation  (23) follows CBH
star product for arbitrary two classical fields $\phi(x),
\chi(x)$ on Minkowski space
\begin{equation}
\phi(x)\,  \star\,  \chi(x) =
\phi(y) \, \hbox{exp}\, i x_{\mu} {\widetilde{\gamma}}\ ^{\mu} 
\left( { \overleftarrow{\partial} \over \partial y} , 
{  \overrightarrow{\partial} 
 \over \partial z} \right) \, \chi(z)\big|_{y=z=x} \, ,
 \end{equation}
 where $ {{\gamma}} ^{\mu} (\alpha,\beta)=
 \alpha^{\mu} + \beta^{\mu} +
  {\widetilde{\gamma}}\ ^{\mu} (\alpha, \beta)$ (see (22)).

 It can be shown that the formula (24) is an example of
Kontsevich formula [23]
 describing star product for arbitrary
Poisson structure, which is written in particular basis.

In the case of solvable algebra (5) the formulae (21) and
(23) can be written in closed form as follows:

\begin{equation}
e^{i k^{\mu} \hat{x}_{\mu} }
\cdot
e^{i l^{\mu} \hat{x}_{\mu} }
     =
     e^{i r^{\mu} (k,l)  \hat{x}_{\mu} }\, ,
     \end{equation}
     where
      $f_{\kappa}(\alpha) = {\kappa\over \alpha}
       \left(1 -  e^{- {\alpha\over\kappa}} \right)$,

\begin{eqnarray}     
r^{0}(k,l) &=& k^{0} + l^{0}\, ,
\cr
r^{i}(k,l)  &= & {
f_{\kappa}(k^{0}) e^{l^{0}\over \kappa} k^{i}
 +
 f_{\kappa}(l^{0})  l^{i}
 \over
 f_{\kappa} \left( k^{0} + l^{0} \right) }\, ,
 \end{eqnarray}
 and
 \begin{equation}
 e^{ik^{\mu} x_{\mu}} 
 \star 
  e^{il^{\mu} x_{\mu}}  =
   e^{ir^{\mu}(k,l) x_{\mu}} \, .
     \end{equation}
     Differentiating (27) twice and putting $k^{\mu}= l^{\mu}=0$
one obtains two relations
\begin{eqnarray}
x_{0} \, \star \, x_{i} & = & x_{0} x_{i} + {i\over 2\kappa}
x_{i}\, ,
\cr\cr
x_{i} \, \star \, x_{0} & = & x_{0} x_{i} - {i\over 2\kappa}
x_{i}\, ,
\end{eqnarray}
in consistency with (5). In general one gets
\begin{equation}
\phi(x) \, \star \, \chi (x) =
\phi \left( {1\over i} {\partial \over \partial k^{\mu} } \right)
\chi \left( {1\over i} {\partial \over \partial l^{\mu} } \right)
e^{ir^{\mu} (k,l)x_{\mu} } \Big|_{k^{\mu}=l^{\mu}=0} \, .
\end{equation}
The quantization obtained by the multiplication of ordered
exponentials (9) describes other reparametrization of the CBH
star product. Indeed, using the formula (see also [24])
\begin{equation}
e^{i k^{\mu}\hat{x}_{\mu} } = 
:e^{i p^{\mu}\hat{x}_{\mu} }:
=
e^{- i p_{0}\hat{x}_{0} }
e^{i\vec{p}\, \vec{x} }
\, ,
\end{equation}
where

\begin{equation}
p_{0} = k_{0}\, , \qquad
p_{i} = {\kappa\over k_{0} }
\left( 1 - e^{- {k_{0}\over \kappa} } \right)k_{i}
 = f_{\kappa} (-k_{0})
 k_{i}\, ,
\end{equation}
one arrives at the formula (11). The corresponding new star
product $\circledast$ takes the form
\begin{equation}
e^{ip^{\mu}x_{\mu} } 
\circledast
e^{i \tilde{p} ^{\mu}x_{\mu} }  =
e^{ i (p^{0} +\tilde{p} ^{0} )x_{0} 
+ i
\left( p^{i} 
e^{- {\tilde{p}_{0} \over \kappa}}
+ \tilde{p} ^{i}  \right)
x_{i} } \, .
\end{equation}
Differentiating twice relation (32) one gets
\begin{eqnarray}
x^{0} \circledast x^{i} &= & x^{0}x^{i}\, ,
\cr
x^{i} \circledast  x^{0} &= & x^{0}x^{i} - {i\over \kappa}
x^{i}\, ,
\end{eqnarray}
and we reproduce again relations (5).

It should be pointed out that the  star product (32) is more
physical, because the parameters $p^{\mu}$ can be interpreted as
the fourmomenta with the addition law derived from the
$\kappa$--deformation of relativistic symmetries [11--15].

\section{Noncommutative translational Invariance}

Let us observe that the $\kappa$--Minkowski space-time
coordinates $\hat{x}_{\mu}$ and noncommutative translations
$\hat{v}_{\mu}$ satisfy the same Lie--algebraic relations (see
(2) and (4)). If we consider the noncommutative fields
 $\Phi(\hat{x},\hat{v}), \chi(\hat{x}, \hat{v})$, these two
commuting algebraic structures can be represented by the following
extension of CBH star product (24):
\begin{eqnarray}
\phi (x,v) \star \chi(x,v) &=& \phi(y,u) 
\\
&&     
\cdot \hbox{exp} \, i\left\{ x_{\mu} \, \widetilde{\gamma}\ ^{\mu} 
\left(
 \overleftarrow{\partial \over \partial y}, 
  \overrightarrow{\partial \over \partial z}
  \right) +
  v_{\mu}\,  \widetilde{\gamma} \ ^{\mu}
  \left(
 \overleftarrow{\partial \over \partial u}, 
  \overrightarrow{\partial \over \partial w}
\right)  \right\}
 \chi(z,w)\Bigg|_{y=z=x \atop u=w=v}\, .
 \nonumber
  \end{eqnarray}
In particular if $\Phi(\hat{x},\hat{v}) \equiv \Phi(\hat{x} +
\hat{v})$ then one can write

\begin{eqnarray}
\phi (x + v) \star \chi(x + v) %&=&
=
 \phi(y) 
\hbox{exp} \, i
\left(
 x_{\mu} +v_{\mu} \right) 
% \cr
%&&
 \  \cdot \, \widetilde{\gamma} \ ^{\mu} 
   \left(
 \overleftarrow{\partial \over \partial y}, 
  \overrightarrow{\partial \over \partial z}
  \right)
  \chi(z)\Bigg|_{y=z=x} \, .
  \end{eqnarray}
If we introduce the notation stressing the nonlocal character of
CBH star product
\begin{equation}
\phi(x)  \star \chi(x) =
\int \D^{4} x_{1} \, \D^{4}x_{2} 
\hat{K}\left(x;x_{1},x_{2}\right)
\phi(x_{1})
\chi(x_{2})\, ,
\end{equation}
then from (35) follows that the 
 noncommutative translations in $\kappa$--Minkowski space imply
the shift of standard space--time coordinates $x_{\mu}$ by
commuting translations $v_{\mu}$ i.e. in the formula (36) one 
should replace
 $\hat{K}(x;x_{1},x_{2}) \to 
\hat{K}(x+v;x_{1},x_{2})$.
We obtain
\begin{eqnarray}
 \int\!\!\!\int \D^4  {\hat{x}}
 \Phi(\hat{x} + \hat{v}) \chi(\hat{x}+\hat{v}) & =& 
 \int \D^{4}x \, \D^{4}x_{1}\, \D^{4}x_{2}
 \hat{K}(x+v; x_{1},x_{2}) \phi(x_{1})\chi(x_{2})
 \cr
 & = &
  \int \D^{4}x \, \D^{4}x_{1}\, \D^{4}x_{2}
 \hat{K}(x; x_{1},x_{2}) \phi(x_{1})\chi(x_{2})
 \cr
& = &
  \int\!\!\!\int \D^4  {\hat{x}} \, \Phi(\hat{x})
  \chi(\hat{x}) \, .
  \end{eqnarray}
 
%%%%%%%%%%%%%%

For the arguments presented above it is essential linear form of
the Poisson structure.
It appears however, that one can extend the $\kappa$--Minkowski
star product in such a way that general transformations of
$\kappa$-deformed Poincar\'{e} group, with quadratic relations
describing commutator of Lorentz group parameters and
noncommutative translations, can be also considered [25].

\section{Final Remarks}

The discussion presented in Sect. 2-4 (action integrals for
complex noncommutative fields, star products, translational
invariance) can be extended to higher local powers, describing
e.g. the interaction Lagrangeans in noncommutative local field
theories  (for $ \lambda \Phi^{4}$ theory see [18]). These
results will be also applied to the description of Abelian and 
nonAbelian
 gauge  theories on $\kappa$--deformed Minkowski space [24].

\end{document}